ECE 792: Senior Project Final Report

Advisers: Professor Kent Chamberlin, Matthew Argall

# Faraday Rotation Ammeter

ECE Student: Jason Sisk

Department of Electrical and Computer Engineering
University of New Hampshire

May 9th, 2018

# Executive Summary:

      The Faraday Rotation Ammeter (FRA) experiment takes advantage of the Faraday Effect in order to measure the current density in a given environment. The Faraday Effect explains a rotation of a polarized electric field as it propagates through a magnetic field. The FRA uses a polarized laser source that is coupled into fiber optic cable, then placed in the desired environment to measure the change of polarization angle caused by the magnetic field. The change in the angle of polarization can be used to then find the strength of the magnetic field and ultimately the level of current that created it. The Faraday Effect is amplified by the Verdet Constant, a material property that describes the number of degrees the angle of polarization rotates for a given medium. The device is intended to measure the current density seen in the Aurora. Estimated current values are expected to be around the lower micro-amps range (10-100µA/m$^2$) for that application. The main goals of the project were to simulate the Faraday Effect on the propagation of the linear polarized light wave, measure the Verdet Constant, re-build the previous design and determine what components of the design if improved or replaced with newer technology could provide the FRA the required sensitivity.

# Table of Contents

# Introduction:

The Faraday rotation ammeter (FRA) is an optical instrument for measuring currents in space plasmas. It uses the Faraday effect on polarized light in optical fibers to measure the curl of the magnetic field and, hence, measure current density. Several designs of the instrument have been created and tested, but new technology promises to increase the sensitivity and resolution of the device. The main goal of the project is to create a physical prototype of the FRA based on a previous design, and demonstrate the Faraday effect to estimate the degree of sensitivity that can be realized with further improvements.

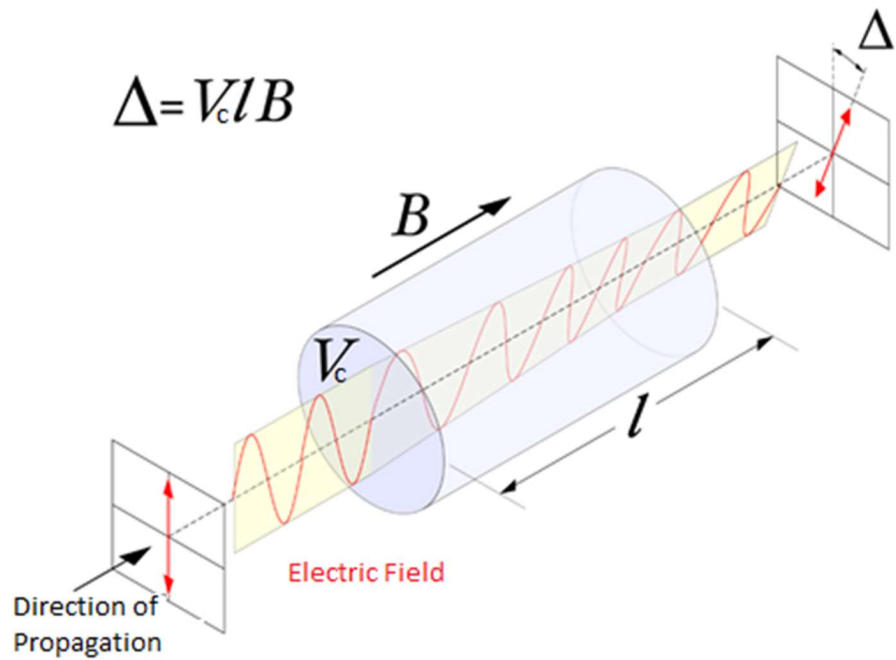

Figure 1.0: Visual Demonstration of Faraday's effect

Figure 1.0 Illustrates a simplified device for measuring the Faraday effect while the full design is explained further in the Approach. The Faraday Effect is seen when the linearly polarized light enters a medium in the presence of a magnetic field (B). The Verdet constant, Vc, is a material property that describes the number of radians the angle of polarization rotates per

tesla per unit length of propagation. This angular rotation is known as Faraday's Effect and the relationship between the change in polarization angle ($\Delta\theta$) and the magnetic field is given by Equation 1 below.

Equation 1: Faraday′s Effect $\Delta\theta = V_c * \oint \hat{B} \cdot \widehat{dl}$.

In Equation 1, Faraday's Effect occurs along a straight path as shown in Figure 1.0 and described above allowing one to measure the strength of the magnetic field. In the FRA1 final design the path is placed around a loop to further increase the sensitivity of the device by increasing the length the medium and allowing one to measure the current density.

This project set three goals on the proposal date; to build the FRA instrument, create a simulation of the Faraday Effect and investigate new technologies. The 1st goal was to identify the optical components required for the FRA, assemble FRA device, and test the FRA to obtain results. The FRA uses the following components: a laser source, a polarizer, half-wave plate, beam splitter, fiber cable, mirror, analyzer circuit and detector. To better understand each piece and for a definition of their purpose see the section on Design/Setup.

The 2nd goal dealt with creating a simulation of the design and how theoretically the linearly polarized light wave is affected as it travels through the FRA. The simulation walks through the same steps as the physical design but the components are represented by functions that portray what would happen as the light wave travels through. Using MATLAB, a magnetic field induced along a fiber ring or waveguide is portrayed using values for a magnetic field and coordinates along a circle or line. Then, the angle of rotation that occurs from the Faraday Effect is used to calculate the Verdet Constant of the medium. Along with that, the model simulates the magnetic field and rotation angle, then provides a measurement of the current density through the use of Ampere's Law.

The 3rd goal was to use the information identified in the modeling process to design a device with greater sensitivity, and an improved design will be created later based upon those findings. The proposed sensitivity of the new design was to be calculated to see if the improvements are sufficient to measure the current levels around the microamps range needed for its application.

The results of the setup and simulation created provided whether or not the design could be further improved to meet the assumed requirements needed to measure the current in the aurora. The results of the physical setup and simulations along with available components was used to propose an estimate of the sensitivity and final cost for a research proposal in the hopes of future funding.

## Background:

The idea of using the Faraday Rotation Ammeter has been studied before in the hopes of using this alternative concept to measure the current in space applications via the Faraday Effect. The Faraday Effect was discovered in 1845 by Michael Faraday, when he noticed that the polarization of light is affected by a magnetic field along the length of the medium as shown in Figure 1.1.

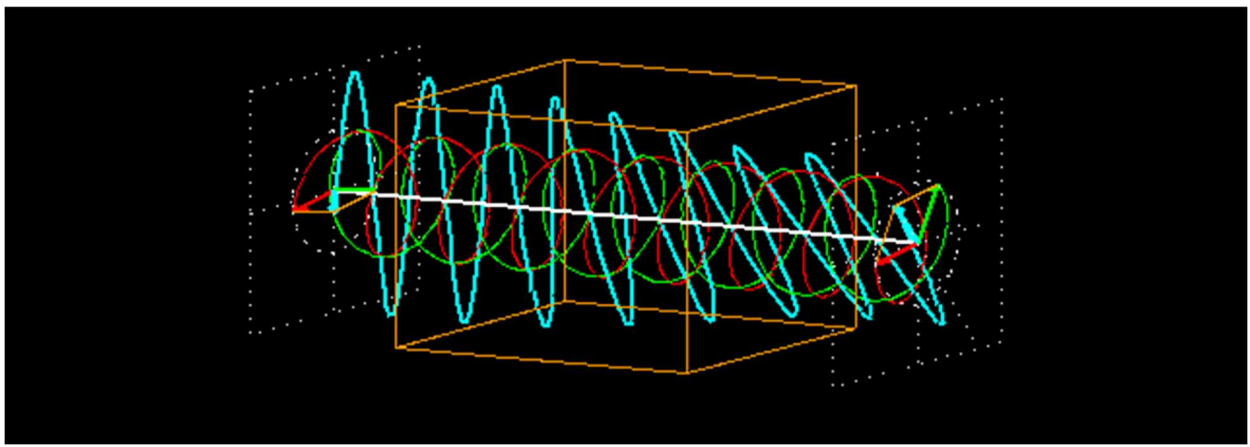

Figure 1.1: Breakdown Visualization of the Faraday Effect

The Faraday Effect can be further explained through a description of the waves that make up the linearly polarized light. In Figure 1.1 the linearly polarized light wave is made up of both a "left-handed" (red) and a "right-handed" (green) wave. When the polarized light wave is propagating through the medium embedded within a magnetic field, both of the wave elements are affected. One of the wave elements lags behind the other based on the direction of the magnetic field resulting in an angle rotation. When the elements are combined they show how the electric field vector (cyan) rotates as it propagates through the medium. The difference in the angle or rotation of the polarized light wave can be found through knowing the Verdet constant of the medium, the length of the medium effected by the magnetic field and the strength of the magnetic field (Equation 1). Once the strength of the magnetic field is found for a loop, the current density that corresponds to the magnetic field strength could be derived through Ampere's Law.

Equation 2: Ampere's Law $\nabla \times B = \mu_0 * J$

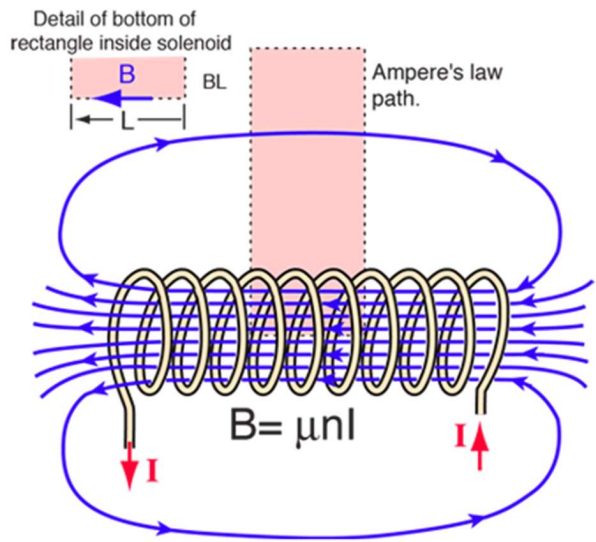

Figure 1.2: Ampere's Law Visualization

25 years ago, two theses were written in collaboration about the Faraday Rotation Ammeter and the results of different designs (Carol L. Strong[1] & Indu Saxena[2]). The intentions for this device is to be used to measure the current density, more directly to be used to measure the currents of the Aurora. Current values are expected to be around the lower micro-amps range (10-100µA/m$^2$). Many different designs were analyzed in order to obtain the best sensitivity and remove the effects of mechanical stress, thermal stress and unwanted birefringence on the system. Through their research a new type of fiber was created to minimalize the linear birefringence and maximize the circular birefringence within the fiber cable itself. This new fiber cable called Spun Elliptically Birefringent fiber (SEB) which is insensitive to mechanical noise and 80% sensitive to the ideal Faraday effect (Carol L. Strong[1]). The difficulties of achieving the sensitivity with the components used in the 90's led to the conclusions that the Faraday Rotation Ammeter method couldn't be used to obtain a small enough sensitivity. The overall goal of this project is to see that if with today's technologies it would be possible to achieve such sensitivity with an up-to-date version of the Faraday Rotation Ammeter.

## Approach:

### Research –

My initial task was to search for resources and familiarize myself with the state of the project as it was 25 years ago. To do this an understanding of Faraday's Effect was obtained through an Optics Lab showing the effect first hand and reading about how it can be used to determine the strength of the magnetic field. Continuing forward, a general understanding of Electromagnetics and Ampere's Law was needed in order to convert the strength of the magnetic field into a current value. A basic understanding of Optics definitions, laws and limitations for each piece within the design (Figure 1.3) was needed to insure proper understanding of how each component works.

A collection of optics equipment along with a table were obtained. Discussion of the readings, equipment found and how to go about the breakdown of the Faraday effect were reviewed during each of the weekly meetings with the help of Matthew Argall and Professor Chamberlin.

Five total designs of the FRA were created but the FRA1 was used for this project due to its simplicity. The other designs can be seen in A.1 with details about their pros and cons. Each piece of the FRA1 design, from the laser source, fiber cabling, beam splitters to the polarizer and detector was analyzed through the known specifications on each component in order to achieve an improved design and to test for its sensitivity. With these performance models, the specifications or variations seen in alternate or new components could be substituted in order to demonstrate how the sensitivity of the overall design would change. Each component's effect on the linearly polarized wave was simulated in MATLAB in order to simulate the physical light wave as it travels through the design. In order to do this, research into how to model the effects of each component on the polarized light using Maxwell's equations in MATLAB was needed.

Design/Setup –

    As stated above, the previous FRA1 design will be used for this project for the simplicity and lower cost that accompanies it.

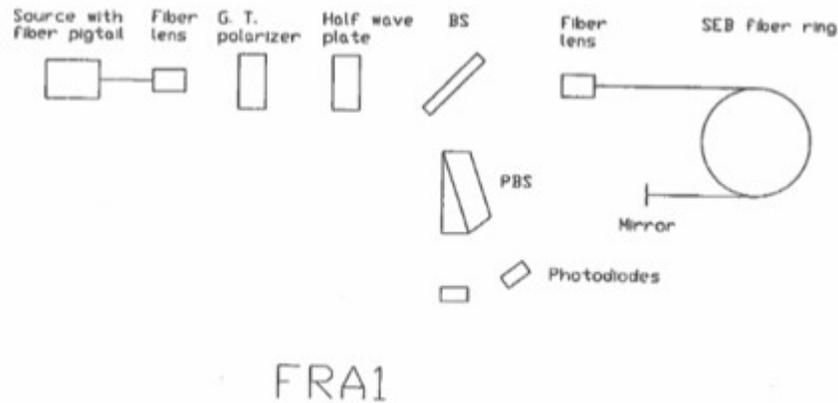

Figure 1.3: FRA Design

The laser source (a Laser Diode) is directed along the fiber into a fiber lens that points the light wave into a polarizer. The polarizer filters the light wave, leaving behind a linearly polarized wave to continue through. The linear polarized light wave continues into a half wave plate where the wave is rotated to prepare the beam to be able to go into the fiber lens and fiber. This half wave plate was chosen based on the laser source wavelength. Once the polarized light wave's plane had been rotated it moves into a beam splitter. The beam splitter splits the light allowing 50% of the light wave to move into the fiber and the other 50% heading off as waste. The light heading into the fiber is transmitted inside the core of the fiber optic cable until the wave reaches a mirror and then returns back through the cable. Along this section of the design a magnetic field is in the direction of propagation in order to cause the plane of polarization to rotate due to the Faraday's Effect. This rotation is seen once it returns to the beam splitter and is reflected down into the photodiodes in the detector circuit and then compared to the calibrated data. This design was setup using an optics bench in order to test and obtain results of the rotation angle. The rotation angle was then used to find the current of the magnetic field that caused the rotation.

## Modeling/Testing -

A simulation was created to show how the wave propagates via MATLAB in order to understand the Faraday Effect. The goal of the simulation was to allow for each component of the design's ideal effect on the light wave to be understood. With each component's general effect applied in MATLAB, the system was simulated showing how the design and effect works. In the simulation, newer components could be added in and if the sensitivity improved would lead to an overall better design can be obtained. This would provide data to prove whether the newer technology of any kind was worth investing in rather than buying the equipment and finding out later when measuring the sensitivity.

# Detailed Design Documentation and Results:

## Gathering Equipment -

The first steps to gathering the equipment were taken through the help of Professor Richard Messner and the Physics Department. A small two by three-foot ThorLabs optics table was acquired through the generosity of Professor Messner and was moved into a project work space in Morse Hall. Multiple other components were found due to the help of Matthew Argall, Professor Roy Torbert, Professor Marc Lessard and the Physics department that were left over from the previous FRA project. The equipment collected was a variety of different components that could be used for the FRA1 design including mounting equipment, beam splitters, photodiodes, polarizers, half/quarter wave plates, mirrors and more. A list of all the relevant equipment was made by documenting their specifications and used to help prepare the instrument.

## Building the Simulation of the Faraday Effect -

Two models have been built to simulate the Faraday effect in order to calculate the Verdet constant of a medium. Each one takes into account a different perspective, one using the wave equations derived using Maxwell's equations and the other using the physical breakdown of the linearly polarized wave to understand how each component on the smallest level effects the other.

To begin creating the models of the effect, the FRA was closer examined by reading the theses written by Saxena and Strong were done to better understand not only the results of the project back in the 1990's but the route they took to obtain those results. A further understanding of the Faraday effect and how to break the effect down into smaller parts was completed using known electromagnetic and physics equations. Starting with a linearly polarized electric field

($Eo * \cos(kz + \Theta)$) and the value for conductivity in a glass medium ($10*e^{-12}$ S/m), the resulting current density was solved for in both the X and Y components. Using the derived current density values, the charge of a free electron (q = $-1.602*10^{-19}$ C) and the derived electron charge volume density the velocities of a charge along the linearly polarized electric field was found. These velocity vectors along with the value for the Magnetic field were then used to find the resulting electric field that adds onto the components of the original electric field. This relationship comes from the Lorentz Force equation.

Equation 3: Lorentz Force $F = qE + q(v \times B)$

This simulation didn't fully work but the hope was that in adding in one of the linearly polarized waves dimensions it would cause one to propagate faster or slower than the other depending on the direction of the magnetic field. Then when the adjusted electric field component value was added to the other it creates the same complete linearly polarized electric field but the rotation from the initial polarization state could be seen. A goal with this was to see when the medium changed from extreme lengths of fiber to smaller lengths of doped glass where the Verdet constant was higher a similar or better sensitivity would be seen.

In the hopes of comparing the model to results from other researchers a paper from the School of Electrical and Electronic Engineering in Beijing China the Verdet Constant was read in order to further break down the Faraday effect (Li, Y[5]). The approach starts with basic electrical and magnetic fields and manipulates them with Maxwell's equations in order to take into account the difference of the index of refraction from the left-handed to the right-handed waves that make up the polarized light wave traveling in the medium. The induced magnetic field is simulated going in parallel to the direction of propagation of the linearly polarized light wave as done in the model above. As the light wave propagates the difference in the index of refraction causes a phase delay in either the right-handed or left-handed waves, depending on the direction of propagation of the Magnetic field, causing a rotation in the polarization angle of the linearly polarized light wave. This change as described throughout the paper was used to then find the Verdet constant of the medium. The current result of this model, when the indexes of refraction were chosen, lead to the correct estimation of the Verdet constant for SF-57 glass at 23 rad/m-T. Plots from this model, Figure 1.4, show how the X and Y components of the linearly polarized wave change over time along with the change in the overall polarization angle of the linearly polarized wave.

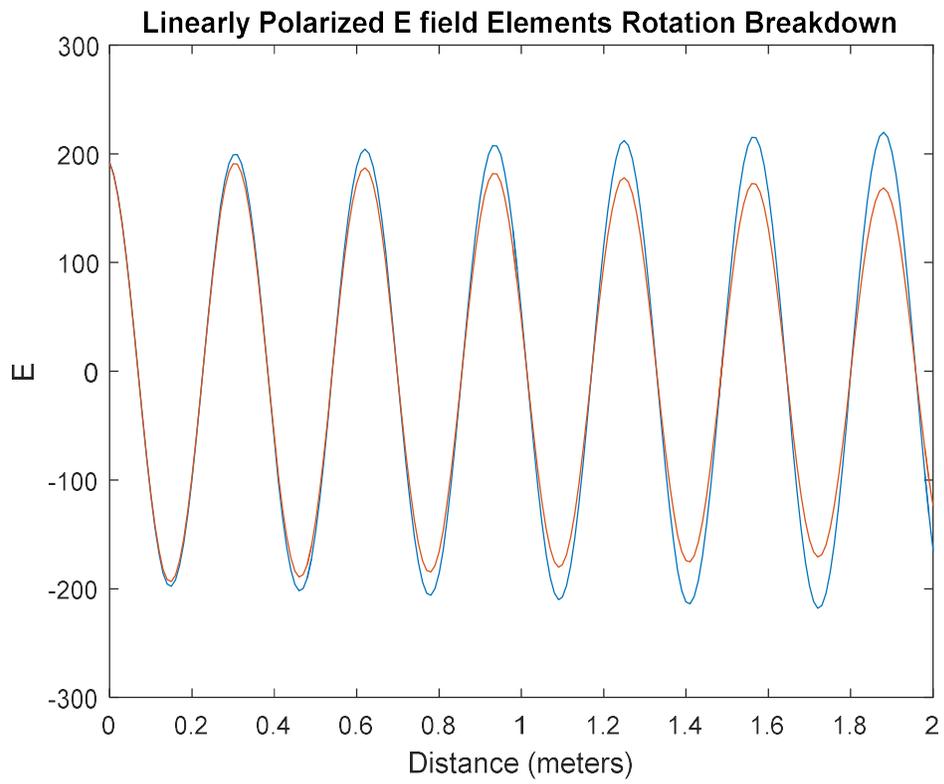
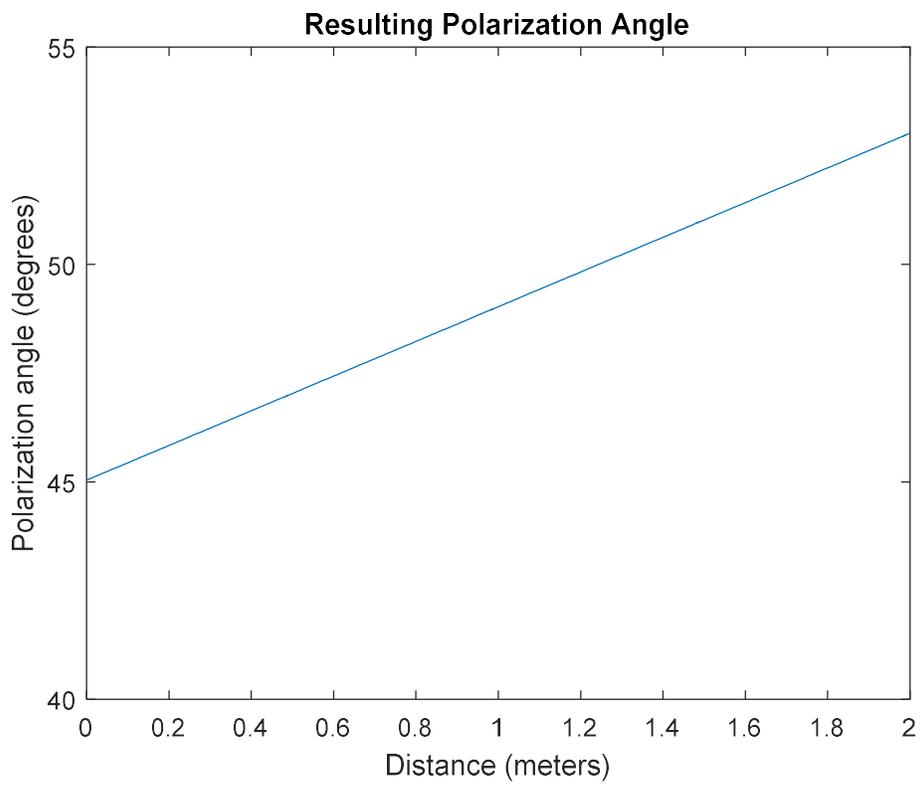

Figure 1.4: Eout and Polarization Angle Plots from Comparemodel.m

In Figure 1.4 above the effect of a constant magnetic field on the linearly polarized wave elements is shown in the Eout plot while the resulting angle of polarization is constantly changing over time (See A.7 for the Matlab code used in both models).

## Measuring the Verdet Constant -

The Physics department gave access to a Faraday Effect Lab (See A.2 for a Picture of the Setup) in order to witness and obtain data using physical equipment. The lab instructions provided an outline of what the Faraday Effect does to a linearly polarized light wave when traveling through SF-57 glass with a constant 30m-T magnetic field. This was set up with a waveguide of the glass surrounded by a solenoid that was not powered, powered to obtain a positive magnetic field and powered the opposite way in order to obtain a negative magnetic field with regards to the direction of propagation. This allowed one to properly test the Faraday Effect as the magnetic field would be observed in three different ways to see the initial state of polarization, a positive change in angle rotation and a negative change in angle rotation. An interesting note is that when the solenoid wasn't powered a slight change in rotation was still seen which was accounted for by the Earth's Magnetic field. Going into the waveguide was a linearly polarized light source with a wavelength of 650nm and a polarizer right before the waveguide. The polarizer was used to find the angle of polarization and the output intensity of the linearly polarized light wave passing through into the waveguide by rotating it from 0 – 360 degrees in five-degree increments. With this and the detector, the point of maximum output was found when the two polarizers matched up as the largest voltage was detected at this point. As expected with Malus's law, when the polarizer was then rotated 90 degrees from this position little to no output voltage was seen at the detector (Equation 4).

Equation 4: Malus's Law  $I = Io * \cos(\Theta)^2$

Malus's Law states that when polarized light is propagating into an analyzer the intensity of the light seen (I) by the analyzer is directly proportional to the square of the cosine of the angle between the transmission axes of the analyzer and the polarizer.

To obtain the results, the polarizer was moved in 10 degree increments while the detector output in mV, seen on the multimeter, was written down. The results were brought into MATLAB in order to discover the rotation of the polarization angle that occurred. The change in angle was calculated with Malus's Law as the incident and resulting wave were measured. With the change in polarization discovered, the solenoid was simulated in MATLAB based on its physical characteristics and the waveguide marked at its length of 0.1m in faradayeffectlabresults.m (See A.6 for the Matlab code). The change in polarization angle was then converted to the Verdet Constant through dividing the angle by the magnetic field strength and the length of the waveguide. With the results from the Faraday Effect Lab a calculation of 28.7 rad/m-T was derived for the Verdet contstant of SF-57 glass (marked as having a Verdet constant of 23 rad/m-T) using the MATLAB curve fitting function and Malus's Law.

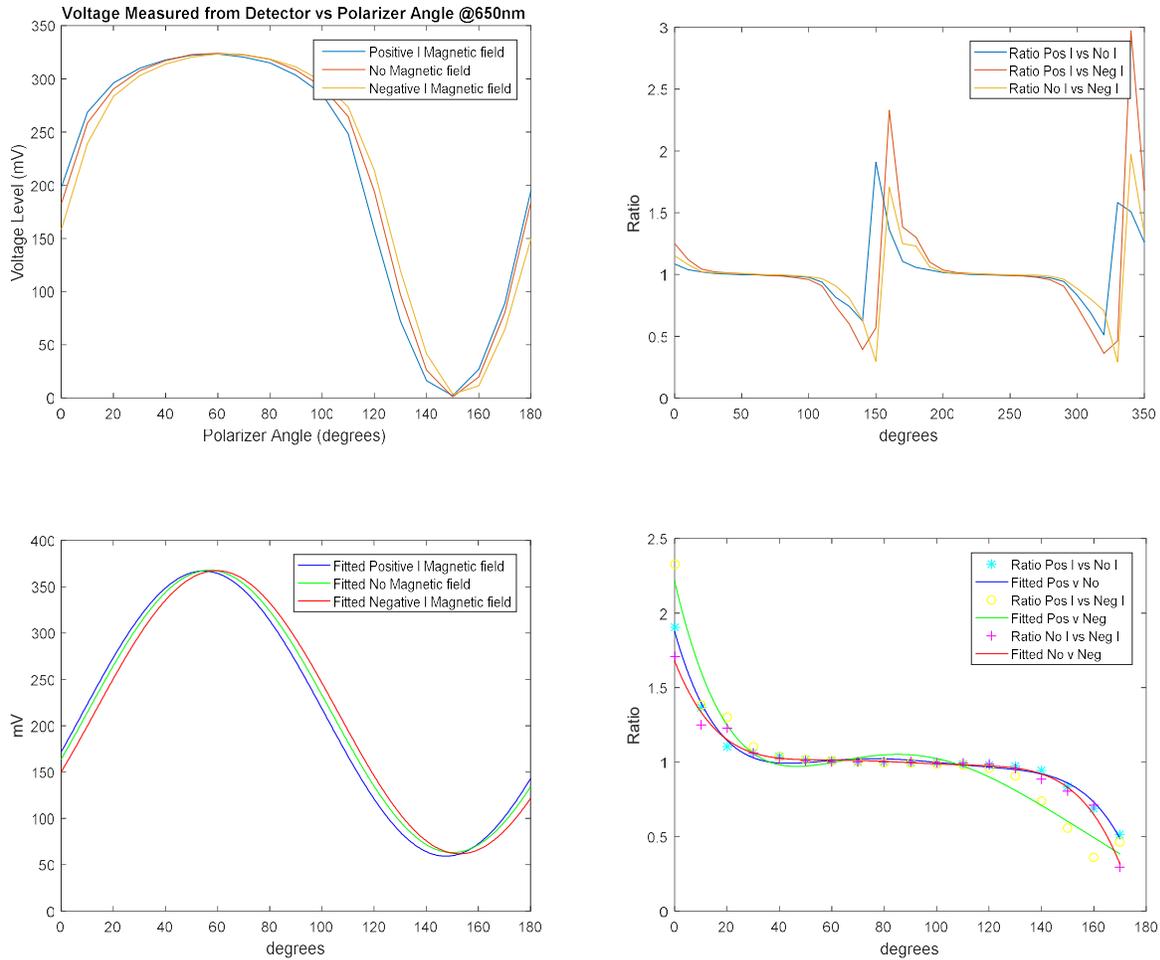

Figure 1.5: Output plots from Faraday Effect Lab, Measured Data and Fitted Data for 650nm. The left plots show the voltage levels of the data side by side. The top left shows the actual measured and the bottom left shows the fitted data through a using a fitting function. The fitting function was used as large ratio values weighed too heavily on the Verdet Constant calculation and to better average out the change in polarization. The ratios between for when no field was present, a magnetic field in the direction of propagation (labeled positive) was present and magnetic field going against the direction of propagation (labeled negative) was present are in the plots on the right. The top right plot shows the ratios left alone and the bottom right shows the ratios with the fitting function applied to the original data. Both plots of ratios appear like inverse tangent functions. The reason the ratios were considered is due to Malus's Law as the ratio of the output over the incident electric field equals to the cosine of the change in the angle of polarization. This data was used to help distinguish the change in polarization as the ratios between these data sets gave rise to the change in intensity when a magnetic field is present in the waveguide. Using the ratio values and Equation 1 above the Verdet constant of the medium was found by comparing the exact value resulting in around a 15-20% difference.

The Verdet Constant is also seen in the following relationship, Equation 5, to change in response to the index of refraction of the medium and the wavelength of the source. This dispersion relationship can be well defined through using the known equation for the medium's index of refraction as well as the gain coefficients for that equation.

Equation 5: Verdet Constant Dispersion Relationship $V = -\frac{e}{2*m*c} \lambda (\frac{dn}{d\lambda})$

Equation 5 shows the Verdet Constant as a function of the source's wavelength and the material's dispersion properties. Based off Equation 5, different wavelength sources were looked into in order to prove that the Verdet Constant would change and to determine how much the Vc would change. An increased Vc would give rise to a greater sensitivity within the instrument. A 532nm HeNe laser was obtained thanks to the Physics Department and then tested as before to calculate the Verdet Constant. Equation 5 predicted that with change in the wavelength the Verdet Constant would increase or decrease. The lab was redone with the new laser source at 532nm and the polarizer rotated in increments of 5 degrees. The results are seen in Figure 1.6. In Figure 1.6, a larger difference in the measured voltage was seen, resulting in an increased Verdet Constant for SF-57 of 37.34 rad/m-T. The predicted value for SF-57 at a wavelength of 532nm is around 38 rad/m-T.

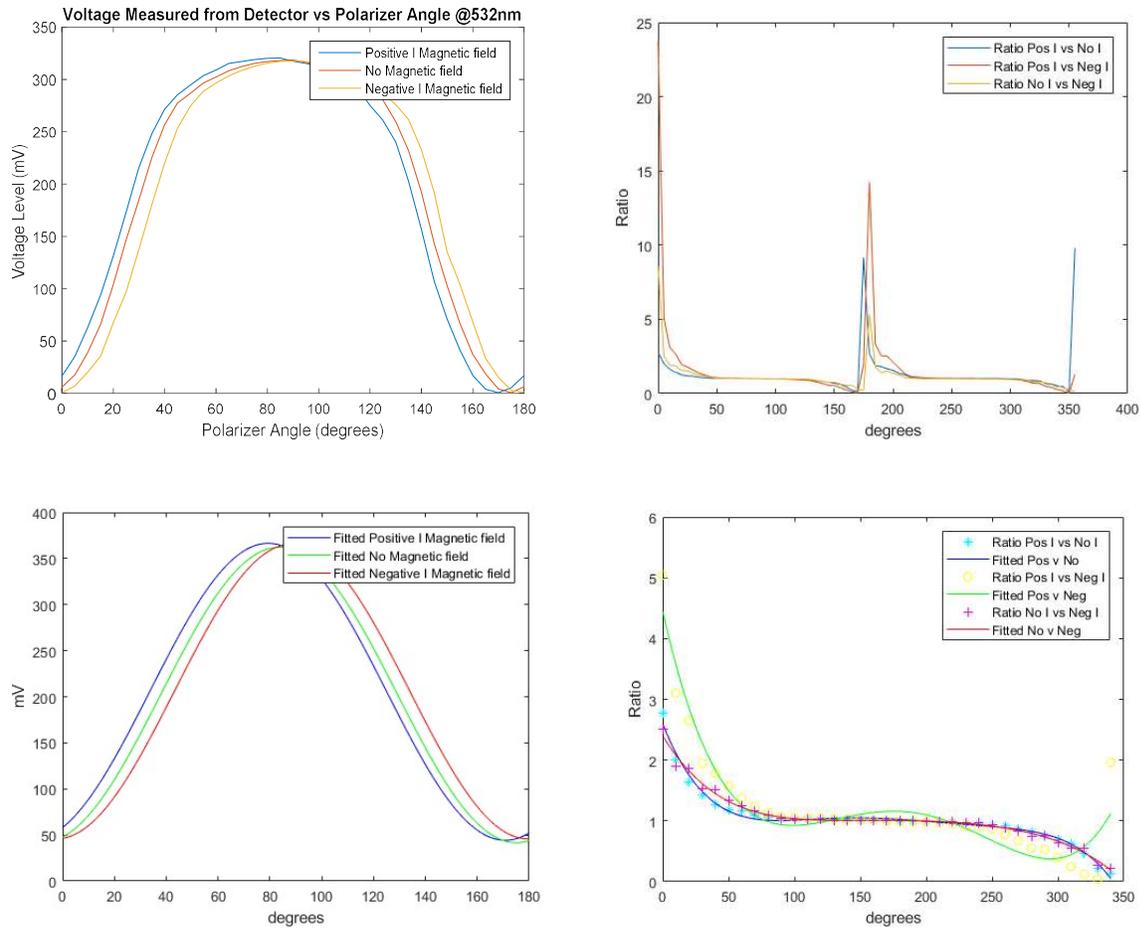

Figure 1.6: Output plots from Faraday Effect Lab, Measured Data and Fitted Data for 532nm. Exactly like Figure 1.5 the left plots show the voltage levels of the data side by side. The top left shows the actual measured and the bottom left shows the fitted data through a using the fitting function. The ratios between for when no field was present, a magnetic field in the direction of propagation was present and magnetic field going against the direction of propagation was present are in the plots on the right. The top right plot shows the ratios left alone and the bottom right shows the ratios with the fitting function applied to original data. In the plots of the measured data (left) that the output voltages are shifted more than before showing that indeed a greater change in the polarization angle was verified before the Verdet Constant was calculated. Taking the experimentally calculated Verdet Constants for both wavelength sources and the dispersion relationship response for most glass (Equation 6), Figure 1.7 shows the theoretical

values compared to the experimental. Alongside the data points for SF-57 is the response of Bulk Fused Silica which is widely used for fiber optic cable and is the type of glass material used in the previous iterations of the FRA.

Equation 6: Sellmeier Dispersion Equation

$$n^2(\lambda) = 1 + \left(\frac{B1 * \lambda^2}{\lambda^2 - C1}\right) + \left(\frac{B2 * \lambda^2}{\lambda^2 - C2}\right) + \left(\frac{B3 * \lambda^2}{\lambda^2 - C3}\right)$$

(See A.5 for B and C values used in dispersion.m)

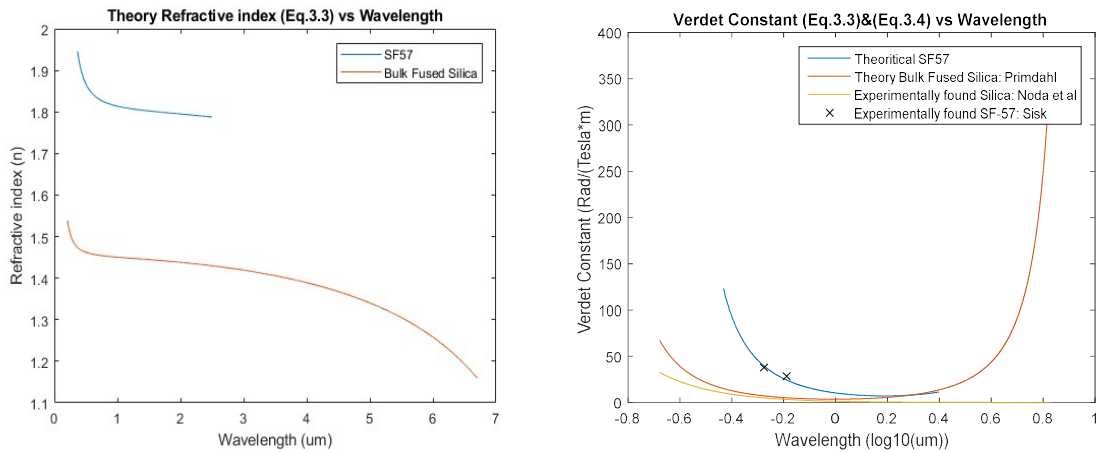

Figure 1.7: Output plots from dispersion.m, Refractive Index and Verdet Constant vs Wavelength. Looking at the projected Verdet Constant at the shorter wavelengths for bulk fused silica and longer wavelengths, we began considering the possibilities of using a different operating wavelength. To see how the wavelength would affect the setup of the device, the equation for Current Sensitivity, Equation 7, was brought forth.

Equation 7: Current Sensitivity $Imin = \frac{\theta min}{(V * Nturns)} = \frac{\pi * r * \theta min}{(V * Leng\ )}$

The aim to use an increased Verdet Constant through greatly increasing or shortening the wavelength produced greater sensitivity in theory. Upon looking further into the attenuation loss versus wavelength for bulk fused silica, the loss in dB/Km greatly increased (30dB/Km at 405nm) for shorter wavelengths due to Rayleigh Scattering and as well for larger wavelengths due to Infrared Absorption (Figure 1.8).

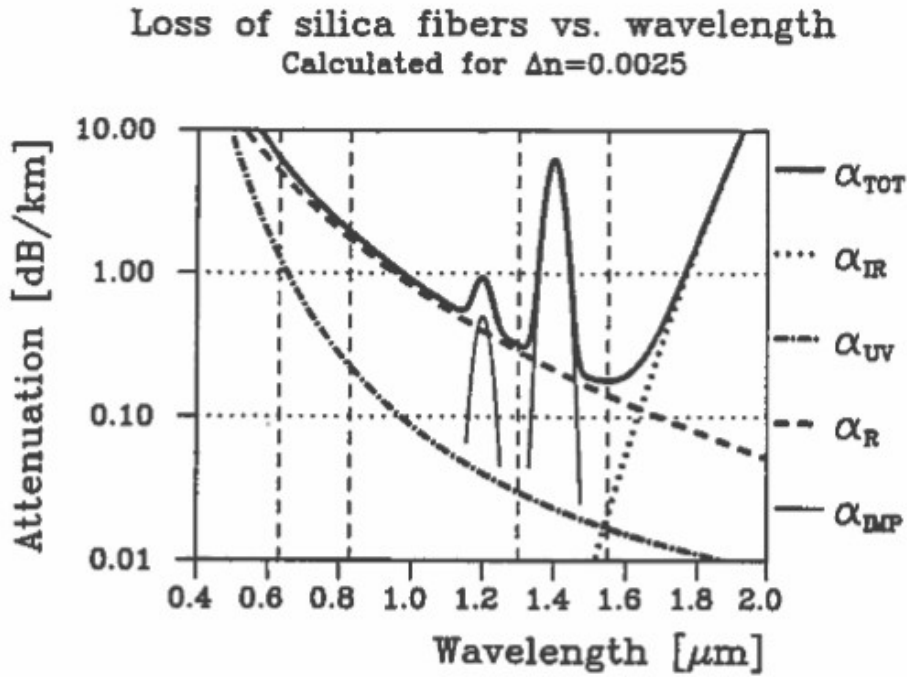

Figure 1.8: Attenuation versus Wavelength for Bulk Fused Silica from Strong ($\alpha_{tot}$)

Taking into account the power level needed to detect the resulting light wave at the detector, shorter and longer wavelengths are pushed off unless a low loss fiber optic cable was found at those frequencies. The chosen current wavelength laser source was 830nm in order to be able to use the other lenses and diode detectors to make the setup as easy and cost efficient as possible.

Building The FRA:
    Building the Faraday Rotation Ammeter requires sufficient knowledge with the safety, use and handling of optic equipment to ensure all components remain usable in the design. The leftover equipment from previous iterations of the FRA project provided a majority of the needed pieces for the FRA-1 design from polarizers, lenses, to mounting equipment and even a fiber coupler. A polarizer has been cleaned as best it could through using the drop and drag technique (See A.3) with the giving of cleaner (Isopropyl Alcohol), lint free cloth, gloves and the proper utensils from the Morse Hall cleaning room on the first floor. There is one laser diode mount that outputs to a

db-9 connector for power and contains a small heatsink to better regulate the temperature of the diode to prevent it from overheating. The diode is soldered onto the wires leading to the connector and has the initial lens removed with a set of lenses to polarize and collimate the light wave for the rest of the design. At this moment the current setup has the ability to detect the laser source using a Near-IR detector card. The detector mount contains both photodiodes, the beam splitter and the differential detector circuit board. The output db-9 connector of the detector mount was analyzed and the two bottom left pins output voltage changed as the input laser polarization was rotated. Multiple spools of fiber have been leftover with one being the SEB fiber optic cable used in Strong's design. In order to couple the linearly polarized light wave into the bare fiber the fiber coupler and positioner will be used. One fiber chunk has been obtained and assured that it will work for the original fiber diameter. The setup has the beam splitter ready to be mounted but it will be bypassed to get a basic measurement first without using it and the mirror maximizing the laser transmission while simplifying the design further.

### Terminating the SEB Fiber -

The original SEB fiber was cut and terminated by hand with the help of Jeff Lepak at the IOL but initial measurements showed improper transmission occurred throughout the fiber most likely due to mishandling while spooling the fiber in the past. The termination process was started by removing the plastic cladding around the glass core using a fiber stripper. A Standard Connector (SC) fiber connector was prepared with an adhesive and the stripped fiber, also covered with adhesive, was fed through. After drying and cleaned off the SC fiber connector with the SEB fiber still had extra fiber sticking out. To prepare the fiber for polishing a sapphire cutter was used to score the fiber core to line the fiber with the ceramic cladding of the connector. Once complete the fiber in the SC connector was polished using an air polish and multiple wet polishes to obtain a polished connector (see A.4 on SC Fiber Termination). The SEB fiber with the SC terminations was tested with a fiber inspector and a fault detector revealing issues in transmission within the fiber cable. Quick measurements of Return Loss and Insertion Loss were taken but the complete measurements across the frequency spectrum will be done in a future extension of the project using a Network Analyzer in order to see if the SEB fiber functions correctly and to see the response at the desired wavelength of 830nm.

### Timeline:

The Timeline of the project was shifted with many of the beginning goals entailing further research. The building of a model to simulate the propagation of the linearly polarized light wave through a similar system has been created. This revised Timeline had an extended expected time needed to correctly simulate and obtain results for possibly new materials for the waveguide, source wavelengths or fiber cable values. The main goal to construct and test the FRA-1 design using the original components wasn't fully completed. In the future, a goal will shift to see if a better detector or fiber can be placed to increase the overall sensitivity of the device.

# Faraday Rotation Ammeter

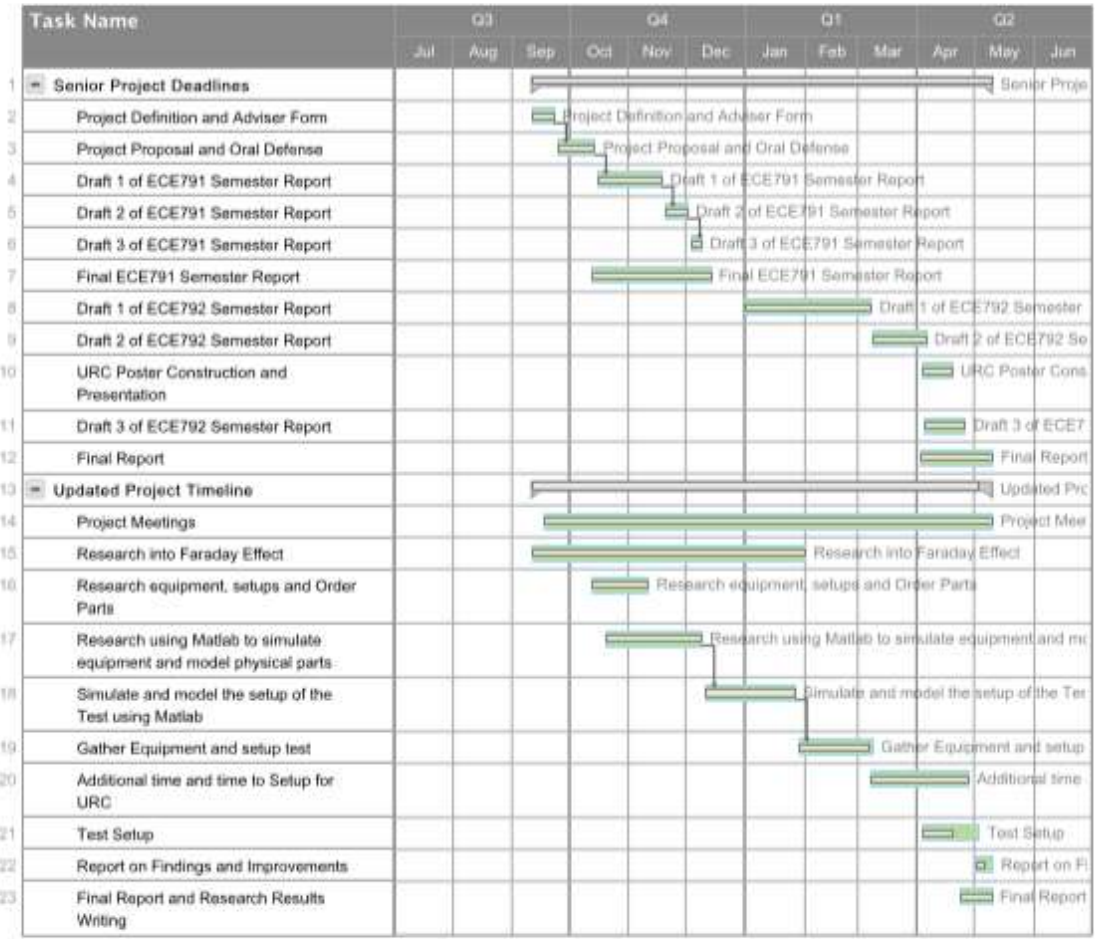

Exported on May 8, 2018 10:18:59 AM EDT

Figure 1.9: Gantt Chart for Project Timeline

The updated Gantt Chart seen in Figure 1.9 shows the Outline of the project. The simulation was complete and the information/equipment has been gathered to be able to fully construct the design. Moving forward the project would be focused on finishing the construction of the FRA-1 and predicting the expected overall sensitivity with the previous performance models.

## Budget:

The budget below is an estimate based off the components seen in the FRA1 design and prices found through Thorlabs, Edmund Optics, Melles Griot and Newport. Each piece below was added to the overall design or used to test the finished system. A majority of the pieces were provided from either the ECE department or the Physics department in order to drive down the total cost. The details and specifications of each component are described in the budget table as well as the expected number of each needed. The budget given by the ECE department was used along with funds allocated by the Physics department.

| Component: | Specifications: | Cost: |
|---|---|---|
| Laser Diode And Current Source | Wavelength: 830nm  632.8 and 830nm found  Power out will be 5mW+ | * |
| Optical Polarizer | Adjustable | * |
| Half Wave Plate | Used to help allow the light to couple into the fiber | * |
| 2 Plate or Cube Beam Splitters | Non-polarizing  Transmission 50:50 | * |
| Fiber Cable (or Appropriate Medium to allow Faraday Effect) | Single Mode SEB fiber  Provided already. | -* |
| Mirror | Silver Painted or silvering the end of the fiber | * |
| 2 Photodiodes | - Used to capture and translate the light into an output current to read | * |
| 2 Fiber Lens or Couplers with Fiber Chunks | Matched to the Laser Source's Wavelength and to the bare Fiber Cable. | $180* |
| Solenoid | - Used to create variable Magnetic field along the Fiber cable | * |
| Power Supply's | - Used to Power the Solenoid, Used to Power the Laser Diode | * |
| Optics Bench | - 2 x 3 foot Obtained from Professor Messner | * |
| All prices were found through Thorlabs, Edmund Optics, or Newport. | *Acquired through UNH or will attempted to be. | Estimated Total: $180  ** Used $90 of ECE budget** |

Figure 2.0: Budget Table

- $100 Budget provided to each ECE student for Senior Project
- Extra budget provided by Physics or ECE department depending on cost
- Based on power output of the expected Laser diode, the instrument will not be turned on at the URC-ISE.

## Ethical Considerations:

This project aims to provide a method of measuring current density using a non-invasive method and doesn't face a moral dilemma with whether or not the project should be continued. The FRA project was created to better the understanding of the natural world and didn't arise from a conflict but from a need to increase the sensitivity of current measurement devices. It does seem morally right to continue the project in order to help push our understanding of the changes in current densities in our atmosphere and possibly beyond.

## Safety:

The Faraday Rotation Ammeter Project uses optics equipment and laser sources that require safety precautions to be followed. The optics equipment must be handled correctly to ensure that the equipment not only doesn't break but stays clean in order to be used to its maximum potential. The classification of laser sources is required to be known in order to prevent irreversible exposure to harmful power levels. Class 3B and class 4 lasers are required to be registered with the Laser Safety Officer (LSO) and have the required signs notifying that a laser is located in the lab and when the laser is in use. A Laser Safety Course was taken and completed along with the Laser being registered with the LSO/Radiation Officer here at UNH. The appropriate Optical Density (OD) rated safety glasses were worn and a basic safe Standard Operating Procedure (SOP) was followed when working with the Laser source.

## Conclusion:

The Faraday Rotation Ammeter project provided a better understanding of the Faraday Effect through research and a simulation, results proving that the Faraday effect is an actual effect that varies with frequency and lastly updated equipment point towards increased sensitivity of the instrument. Through modeling/simulating the design in MATLAB a breakdown of how the components of the light wave are affected were proven. The Faraday Effect and Verdet Constant of a medium were measured using two different laser sources proving that not only does the effect exist but that an increase Vc is seen at higher frequencies which can help improve the sensitivity of the FRA. While researching the design and setup of the FRA, increased sensitivity and reduced noise among the components would provide the device with greater sensitivity. This project although wasn't able to measure the sensitivity of the original device provided insight into new ways of improving the sensitivity through adjusting the device for different wavelengths and working to maximize the Verdet Constant of the medium.

## Future Improvements:

The project left off in the process of waiting for results from a complete setup of the previous FRA1 design. Right now, measurements suggest that choosing a smaller wavelength or higher frequency laser source will provide a higher Verdet Constant in the medium. Increasing the Verdet Constant in the material of the Fiber Cable and the length of the cable is a possible option or avenue needed to be further explored as the fiber is a very important piece. The capabilities of the photodiodes and the possible sensitivity with them needs to continue in order to obtain a better sensitivity. Altogether the main components that effect the rotation of the polarized source wave need to be adjusted to further increase the sensitivity to the desired values of microamps.

## Acknowledgements:

Huge thanks to Matthew Argall and Professor Chamberlin throughout the course of Senior Project. The project wouldn't have gone as far as it did without their questions and guidance.

Special thanks to the following for their help and input on the Project:

Professor Richard Messner, Professor Allen Drake, Professor Olof Echt, Roy Torbert, Jeff Lepak, Ivan Dors, Stan Ellis, and Todd Jones.

# Appendix:
## A.1 Other FRA Designs and Summary

| Design | Thermal Removed | Mechanical Removed | Alignment Difficulty | Sensitivity $A \cdot T/Hz^{\frac{1}{2}}$ | Comments |
|---|---|---|---|---|---|
| FRA1 CW | X | X | Low | 0.5 | SEB fiber<br>Reflect from silvered mirror<br>Surface reflections |
| FRA1 pulsed | X | X | Low | 50 | SEB fiber<br>Reflect from silvered mirror<br>Folded optics, long optical lever arm |
| FRA1.5 |  | X | Medium | 1.0 | Power balance critical<br>No SEB fiber |
| FRA2 |  | X | Medium | NA | One path noisy<br>Pulse instability<br>Folded optics, long optical lever arm<br>No SEB fiber<br>AOM polarization sensitive |
| FRA2.75 |  | X | Medium | 140 | Pulse instability<br>Folded optics, long optical lever arm<br>No SEB fiber<br>AOM polarization sensitive |
| FRA3 | X | X | High | NA | SEB fiber<br>Long optical lever arms<br>AOM polarization sensitive |

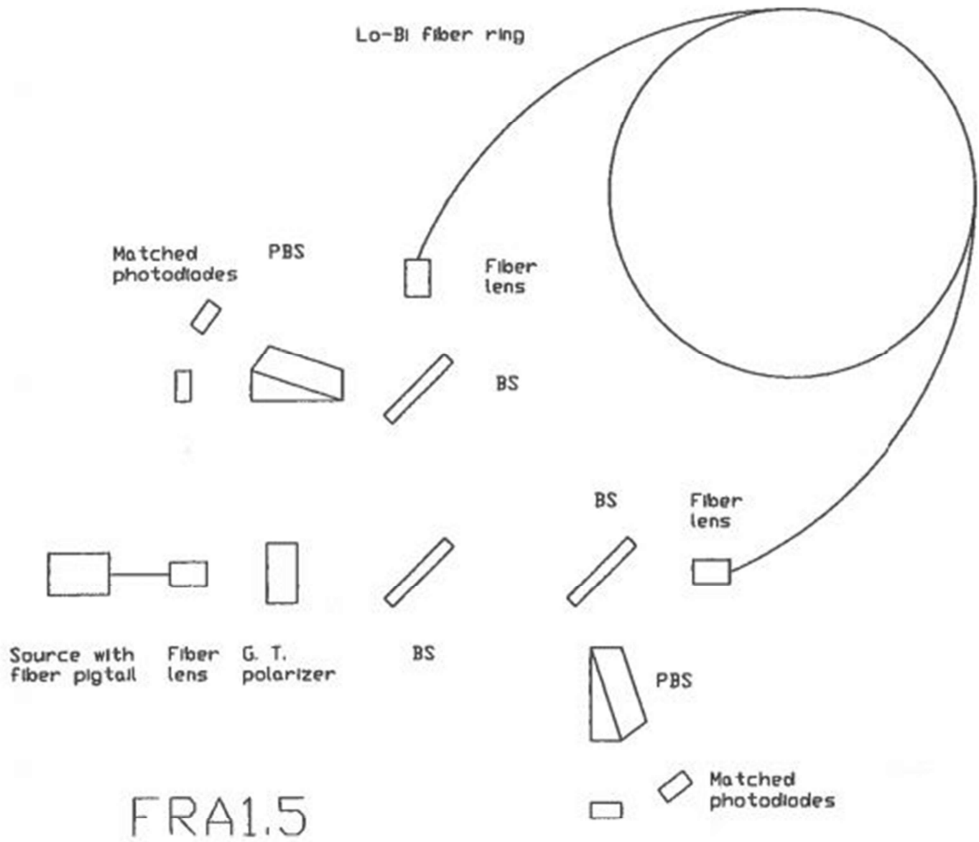

FRA1.5

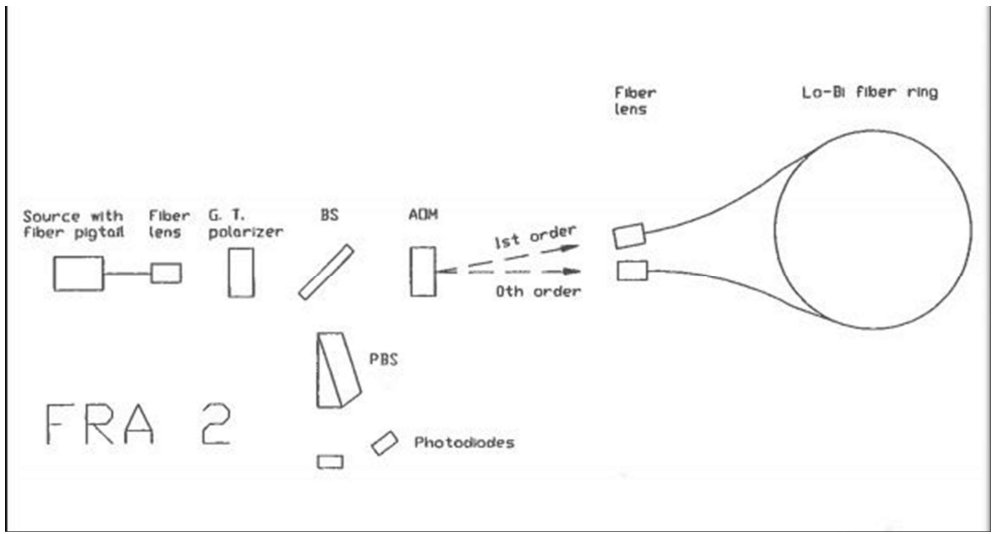

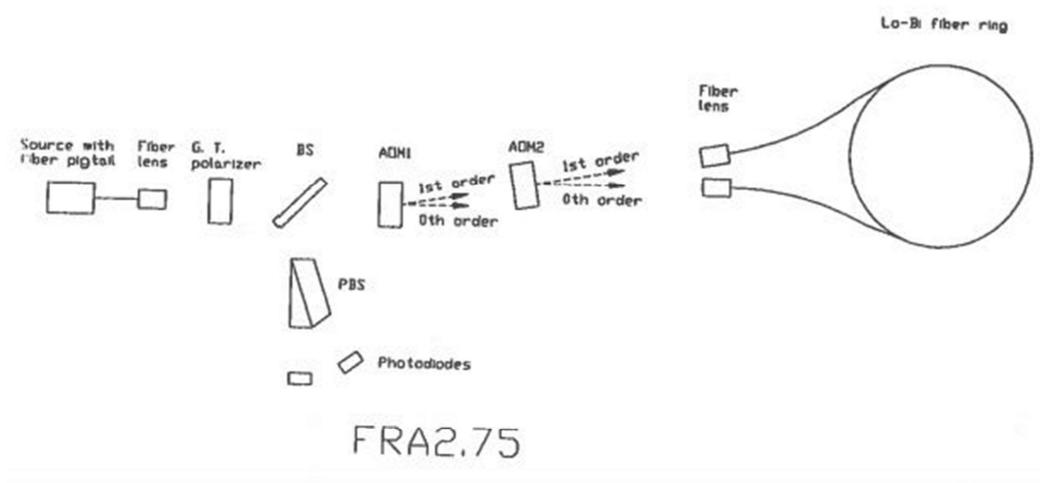

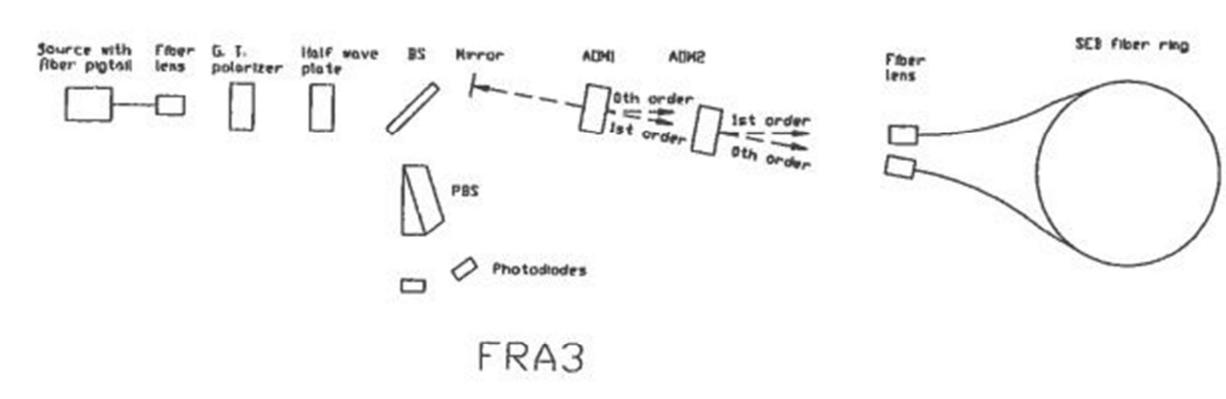

## A.2 Faraday Effect Lab Experiment

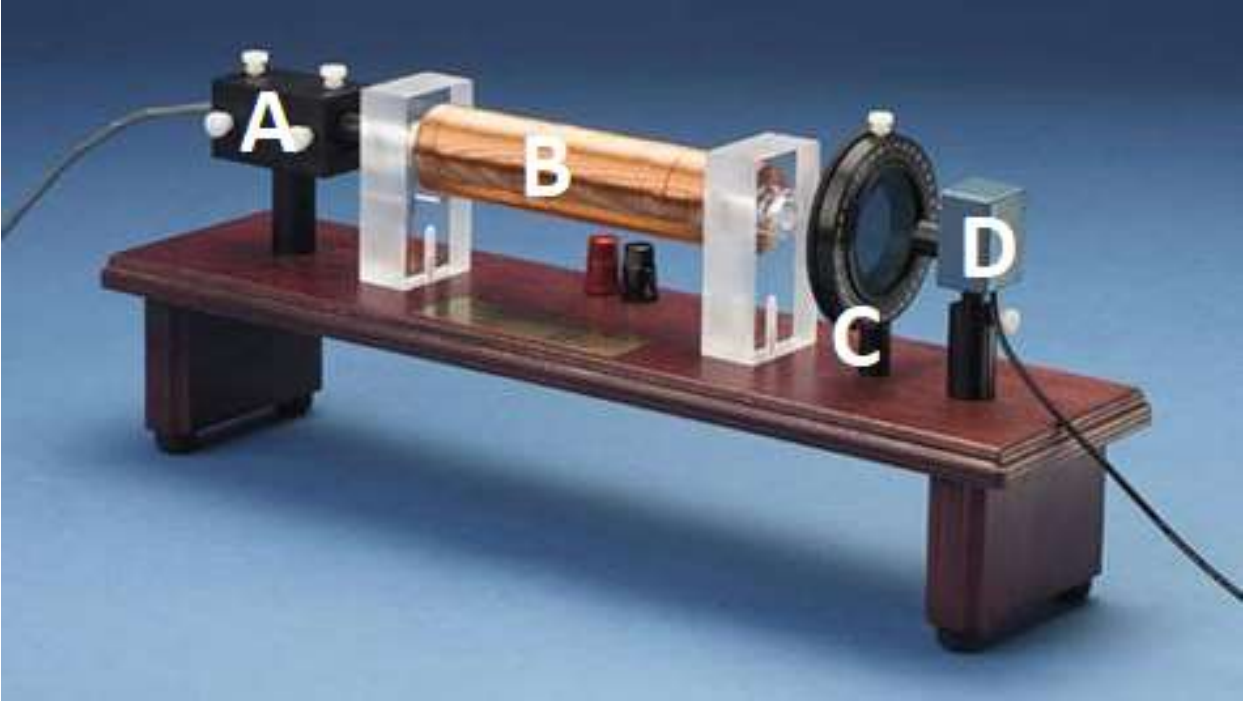

## A.3 Optics Cleaning Procedures

**MELLES GRIOT**  Cleaning Procedures

### Cleaning Methods

**Blowing Method:**

Avoid touching the surface of the optic with unprotected fingers! Always wear powder-free gloves of finger cots.

The first step in cleaning any kind of optic is to remove dust or loosely held particles by blowing them off the surface of the mirror or lens using a dust free blower (a dry nitrogen or $CO_2$ blower will work fine for this). This reduces the chance of scratching the optic during the actual cleaning.

**NOTE:** This is the only cleaning method allowed for bare metal, soft coatings, and pellicles. These types of surfaces should <u>never</u> be touched.

**Drop and Drag Method:**

For this procedure the optic will have to be removed from any mount it is in, and placed in a clean room environment.

1. Use the Blowing Method first.
2. Place a drop of methanol on the side of a lens tissue paper. With the other hand, hold the optic by its edges.
3. Gently place the wet area of the tissue onto the optic.
4. Once the lens tissue is on the optic, slowly drag it across the surface until the lens tissue is dry.

**CAUTION:** Be careful not to drag a piece of dirt across the surface. This could scratch the surface. This procedure should be practiced a couple of times so that streaks are kept to a minimum.

**Wipe Method:**

If the optic cannot be taken out of its held position or if stronger cleaning is required, then the next step will be to wipe using lens tissue paper along with methanol and acetone.

1. Use the Blowing Method first.
2. Fold a piece of lens tissue paper to create a folded edge that is a little longer than the size of the optic. A hemostat is particularly suited to securely hold the folded tissue.
3. Wet the folded edge of the tissue with acetone.
4. Wipe the optic with the lens tissue paper with one continuous motion. Apply some pressure on the wipe to remove stubborn stains.

**NOTE:** Repeat this process, always with new lens tissue paper to eliminate depositing any contamination on the optic. A <u>final wipe</u> with methanol is recommended since methanol does not leave streaks on the surface like acetone. Isopropyl alcohol is also effective, but it too can leave streaks like acetone. If the size of the optic is very small cotton tips can be used instead of lens tissue paper, and the procedure is still the same.

## A.4 SC Fiber Termination

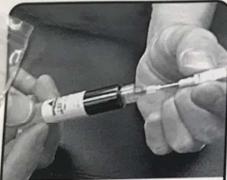

# SC/ST Multimode & Singlemode Connector Termination Instructions

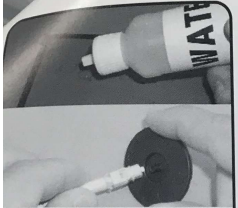

**20** Add a minimum of 3 or 4 drops of distilled/deionized water onto polishing film. Carefully insert the connector ferrule into the ST/SC compatible polishing puck and gently place on pad. Avoid bumping ferule tip on puck or crushing exposed fiber onto pad.

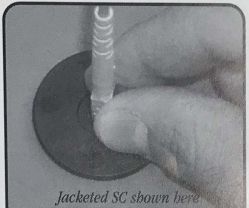

*Jacketed SC shown here*

**21a** Gently grip the connector and apply medium pressure and polish in a 50-75mm [2-3 in.] "figure 8" pattern for 25 to 30 revolutions.

*Important: Do not over-polish and do not use excessive pressure.*
*See Step 21b for efficient use of polishing film.*

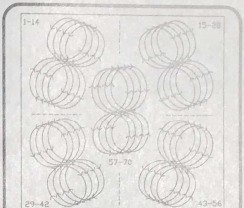

**21b** Applicable to films 2 & 3 only.
*Note: To optimize optical performance of the connectors while maximizing polishing film life, use separate sections of the film per 14 connectors. Using five sections of the film assures a life of at least 70 connectors per film. Variables such as amount of adhesive on tip, size of "figure 8", and polishing pressure can also affect film life.*

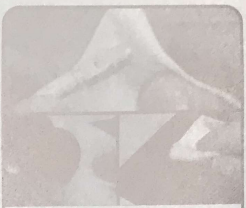

**22 IMPORTANT:** Remove the connector from the polishing puck and clean the ferrule and puck using a lint-free wipe moistened with 99% reagent grade isopropyl alcohol or alcohol-soaked pads. It is also important to thoroughly rinse surface of film with distilled/deionized water prior to storing to assure ideal conditions for next connector.

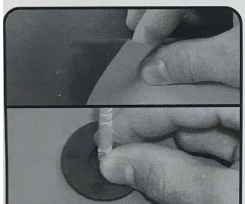

**23 #3 Film (Purple):** Replace the #2 film with the purple #3 film and repeat steps 20-22.

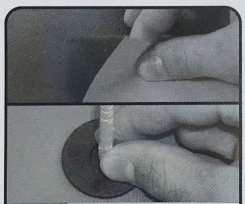

**24a #4 Finishing Film (White):** Required for Singlemode and recommended for multimode especially 50/125 laser optimized applications.
Replace the #3 film with the white #4 finishing film and repeat steps 20 – 22* but use light pressure for 25-to-35 cycles.
*Use step 24b in place of 21b for efficient use of the finishing film.*
*See 24b before proceeding.*

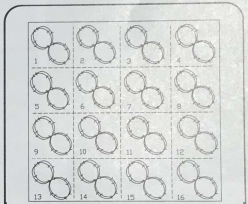

**24b #4 Finishing Film (White):** Use one section of the film per connectors. Be sure to limit the size of the "Figure 8's" to 1.5 inches in height. The same section of this film is not reusable as with the #2 & 3 films.

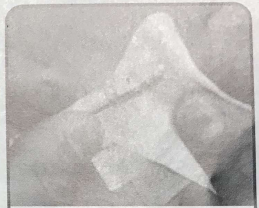

**25** Prior to viewing endface of connector with microscope clean with a dry lint-free wipe.

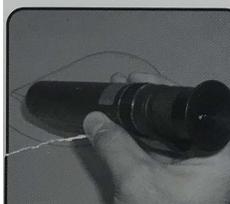

**26** View the polished ferrule surface in the microscope to ensure there are no scratches, voids or chips in the fiber. For proper use of microscope, reference the instructions included with the scope. If polish is acceptable, place dust cap on connector.
*Note: A damaged fiber will result in a high-loss component because it will scatter the light transmission. It is also recommended to check insertion loss and/or back reflection with a power meter and light source.*

Multimode **GOOD POLISH** (XGLO)

Singlemode **GOOD POLISH** (Lightsystem & XGLO)

Multimode **ACCEPTABLE POLISH** (Lightsystem only)
*(Light scratches, dark ring is caused by adhesive)*

**FRACTURE NOT ACCEPTABLE** (Use recovery polish)

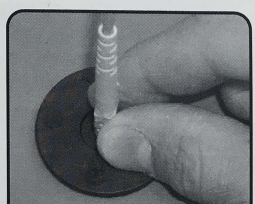

**27 Recovery Polish (if required)**
Scratch Recovery: Repeat steps 23 and 24 (#3 film and #4 film).
Fracture Recovery: This procedure requires the 6 micron recovery film (p/n: FT-PF6) sold separately. Using the 6 micron recovery film, repeat steps 20-22 using medium to hard pressure then start over from step 19 with #2 film.

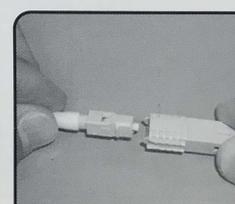

**28 SC Only:** Snap the SC connector fully into the housing. Orient the chamfered corners of the connector relative to the key on the housing as shown.

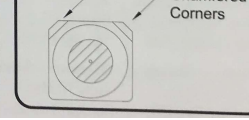

Chamfered Corners

## A.5 dispersion.m

```matlab
%% Dispersion, Verdet and Sensitivity and refractive index against wavelength
% Jason Sisk

%% Explanation

% Saxena's (3.3 and 3.4) and Strong (3.2 and 3.3) Theses both reference the
% equations used below for finding the Verdet constant and current
% sensitivity.
% the materials used are SF57 glass and Bulk fused Silica
% used values seen in both theses and at refractiveindex.info

close all;
%%
% Varying wavelength
wl = 0.37:0.001:2.5; % micro meters
w2 = 0.21:0.001:6.7;

% setting up the arrays
n = zeros(length(wl),1);
n2 = zeros(length(w2),1);

% Vexp = (1.8*10^(-6))*(1/((w2)^2));

% for loops to iterate to calculate index of refraction from equations at
% refractiveindex.info website. link in box
for i = 1:length(wl)
n(i) = sqrt(((1.81651371*wl(i).^2)/((wl(i).^2)-
0.0143704198))+((0.428893641*wl(i).^2)/((wl(i).^2)-
0.0592801172))+((1.07186278*wl(i).^2)/((wl(i).^2)-121.419942))+1);
end
for i = 1:length(w2)
n2(i) = sqrt(((0.6961663*w2(i).^2)/((w2(i).^2)-
(0.0684043^2)))+((0.4079426*w2(i).^2)/((w2(i).^2)-
(0.1162414^2)))+((0.8974794*w2(i).^2)/((w2(i).^2)-(9.896161^2)))+1);
end

% plotting
figure;
plot(wl,n)
hold on
plot(w2,n2)
title('Theory Refractive index (Eq.3.3) vs Wavelength')
xlabel('Wavelength (um)')
ylabel('Refractive index (n)')
legend('SF57','Bulk Fused Silica')

%%

calc650 = 28.67; % from calculations
place650 = 0.650;
calc532 = 37.34;
place532 = 0.532;
```

```matlab
%Constants
m = 9.109*10^(-31); % electron mass Kg
e = 1.60217662*10^(-19); % electron charge coulombs
c = 299792458; % speed of light m/s
thetamin = 7.8*10^(-5); % detectability of photodiode circuit according to
Saxena
Strongthetamin = 0.5*10^(-6);
NturnsStrong = 635;
Nturns = 680; % also seen in Saxena thesis could do (2*pi*r)/L instead

% SF57 dispersion numbers
b1 = 1.81651371;
b2 = 0.0143704198;
b3 = 0.42889341;
b4 = 0.0592801172;
b5 = 1.07186278;
b6 = 121.419942;

% fused silica dispersion numbers
a1 = 0.6961663;
a2 = 0.0684043;
a3 = 0.4079426;
a4 = 0.1162414;
a5 = 0.8974794;
a6 = 9.896161;

% Derived Verdet equation for fused silica and SF57
V1 = zeros(length(wl),1);
for i = 1:length(wl)
V1(i) = (e/(2*m*c))*((wl(i)^2)/(n(i)))*(((b1*(b2))/(((wl(i)^2)-
(b2))^2))+((b3*(b4))/(((wl(i)^2)-(b4))^2))+((b5*(b6))/(((wl(i)^2)-(b6))^2)));
end
V2 = zeros(length(w2),1);
Vexp = zeros(length(w2),1);
for i = 1:length(w2)
V2(i) = (e/(2*m*c))*((w2(i)^2)/(n2(i)))*(((a1*(a2^2))/(((w2(i)^2)-
(a2^2))^2))+((a3*(a4^2))/(((w2(i)^2)-(a4^2))^2))+((a5*(a6^2))/(((w2(i)^2)-
(a6^2))^2)));
Vexp(i) = (1.8*10^(-6))*(1/((w2(i))^2));
end

%Tesla to Amp/m conversion
T2A = 797700;

figure;
plot(log10(wl),V1)
hold on
plot(log10(w2),V2)
hold on
plot(log10(w2),Vexp*T2A)
hold on
plot(log10(place650),calc650,'xk')
hold on
plot(log10(place532),calc532,'xk')
title('Verdet Constant (Eq.3.3)&(Eq.3.4) vs Wavelength')
```

```matlab
xlabel('Wavelength (log10(um))')
ylabel('Verdet Constant (Rad/(Tesla*m))')
legend('Theoritical SF57','Theory Bulk Fused Silica: 
Primdahl','Experimentally found Silica: Noda et al','Experimentally found SF-
57: Sisk')

% figure;
% plot(w2,Vexp)
% title('Exp Verdet Constant (Eq.3.4) vs Wavelength')
% xlabel('Wavelength (um)')
% ylabel('Verdet Constant (Rad/(Amp)')
% legend('Bulk Fused Silica')

%% Sensitivity

% Setting up arrays
Bsense1 = zeros(length(wl),1);
Tsense1 = zeros(length(wl),1);
Bsense2 = zeros(length(w2),1);
Tsense2 = zeros(length(w2),1);
Tstrong2 = zeros(length(w2),1);
Isense2 = zeros(length(w2),1);
Bstrong2 = zeros(length(w2),1);

% other eq for sensitivity: Imin = (2*pi*r*thetamin)/(L*Verdet)
% Calculating the Current Sensitivity
for i = 1:length(wl)
Tsense1(i) = thetamin/(V1(i)*Nturns);
Bsense1(i) = Tsense1(i)*T2A;
end
for i = 1:length(w2)
Tsense2(i) = thetamin/(V2(i)*Nturns);
Bsense2(i) = Tsense2(i)*T2A;
Tstrong2(i) = Strongthetamin/(V2(i)*NturnsStrong);
Bstrong2(i) = Tstrong2(i)*T2A;
Isense2(i) = thetamin/(Vexp(i)*Nturns);
end

figure;
% plot(wl,Bsense1)
% hold on
plot(w2,Bsense2)
hold on
plot(w2,Bstrong2)
title('Theory Current Sensitivity (Eq.3.3) vs Wavelength')
xlabel('Wavelength (um)')
ylabel('Current (A) ')
legend('Bulk Fused Silica Saxena','Bulk Fused Silica Strong') %'SF-57'

figure;
plot(w2,Isense2)
title(' Exp Current Sensitivity (Eq.3.4) vs Wavelength')
xlabel('Wavelength (um)')
ylabel('Current (A) ')
legend('Bulk Fused Silica')
```

## A.6 Faraday Effect Lab Code

```matlab
%% Senior Project FaradayEffectLabResults.m

results = importdata('data - Faraday Rotation Lab.txt');
lensangle = results.data(:,1);
poscurrent = results.data(:,2);
nofield = results.data(:,3);
negcurrent = results.data(:,4);

figure;
plot(lensangle, poscurrent)
hold on;
plot(lensangle, nofield)
hold on;
plot(lensangle, negcurrent)
hold on;
legend('Positive I Magnetic field','No Magnetic field', 'Negative I Magnetic field','Location','NorthEast')
xlabel('degrees')
ylabel('mV')
axis ([0 180 0 350])

posno = results.data(:,6);
posneg = results.data(:,7);
noneg = results.data(:,8);

figure;
plot(lensangle,posno)
hold on;
plot(lensangle,posneg)
hold on;
plot(lensangle,noneg)
legend('Diff Pos I vs No I','Diff Pos I vs Neg I', 'Diff No I vs Neg I','Location','NorthEast')
xlabel('degrees')
ylabel('DiffmV')
hold on;

%% FARADAY ROTATION LAB Verdet Constant Calculations verdetcalclab.m

faradayeffectlabresults

% No fitting method. Calculate the ratios between the data sets and plot
posnofract = (poscurrent./nofield);
posnegfract = (poscurrent./negcurrent);
nonegfract = (nofield./negcurrent);

figure;
plot(lensangle,posnofract)
hold on;
plot(lensangle,posnegfract)
hold on;
plot(lensangle,nonegfract)
```

```matlab
legend('Ratio Pos I vs No I','Ratio Pos I vs Neg I', 'Ratio No I vs Neg I','Location','NorthEast')
xlabel('degrees')
ylabel('Ratio')
hold on;

istart = 2;
istop = 36;
% istart = 17;
% istop = 34;
idiff = istop-istart;

posnoseq = zeros(1,idiff);
posnegseq = zeros(1,idiff);
nonegseq = zeros(1,idiff);
lensangleseq = zeros(1,idiff);

posnoav = 0;
for i=istart-1:istop-1
 posnoav = abs(posnofract(i))+posnoav;
 k = i-(istart-2);
 posnoseq(k) = posnofract(i);
 lensangleseq(k) = (k-1)*10;
end
posnoav = posnoav/idiff;

posnegav = 0;
for i=istart:istop
 posnegav = abs(posnegfract(i))+posnegav;
 k = i-(istart-1);
 posnegseq(k) = posnegfract(i);
end
posnegav = posnegav/idiff;

nonegav = 0;
for i=istart:istop
 nonegav = abs(nonegfract(i))+nonegav;
 k = i-(istart-1);
 nonegseq(k) = nonegfract(i);
end
nonegav = nonegav/idiff;

% Values of the average ratio over range of idiff
val1 = posnoav;
val2 = posnegav;
val3 = nonegav;

%solenoid variables
lsole = 0.15; %length of solenoid
nturns = 1400;
u0 = 4*pi*1e-7;
B = (u0*nturns)/lsole; %(Teslas/Amps)

blabpos = B * 2.6; %magnetic field strength in teslas
blabneg = B * -2.67;
```

```matlab
lengthwg = 0.1; %length of waveguide meters

%Using Malus Law E/E0 = cos(theta)
%For pos current compared to no field
fracposno = (val1);
radposno = acos(fracposno);

verdvalcalcpos = abs(radposno/(blabpos*lengthwg))

%For pos current compared to neg current
fracposneg = (val2);
radposneg = acos(fracposneg);

verdvalcalcposneg = abs(radposneg/((abs(blabneg)+abs(blabpos))*lengthwg))

%For no field compared to neg current
fracnoneg = (val3);
radnoneg = acos(fracnoneg);

verdvalcalcneg = abs(radnoneg/(blabneg*lengthwg))

%% Fitted function results

%creating fitted functions for Voltage seen at detector and plotting them
cfpos = fit(lensangle, poscurrent,'Fourier1','Normalize','on');
cfno = fit(lensangle, nofield,'Fourier1','Normalize','on');
cfneg = fit(lensangle, negcurrent,'Fourier1','Normalize','on');

figure;
plot(cfpos,'b')
hold on;
plot(cfno,'g')
hold on;
plot(cfneg,'r')
hold on;
legend('Fitted Positive I Magnetic field','Fitted No Magnetic field', 'Fitted Negative I Magnetic field','Location','NorthEast')
xlabel('degrees')
ylabel('mV')
axis ([0 180 0 400])

%initializing arrays and variables for the for loop
fitposnofract = zeros(1,180);
fitposnegfract = zeros(1,180);
fitnonegfract = zeros(1,180);
fitradposno = zeros(1,180);
fitradposneg = zeros(1,180);
fitradnoneg = zeros(1,180);

fitval1 =0;
fitval2 =0;
fitval3 =0;
fitidx1 =0;
fitidx2 =0;
fitidx3 =0;
```

```matlab
for q=1:180
    rightindex = q-1;
    leftindex = q;
    
    fitposnofract(leftindex) = (cfpos(rightindex)/cfno(rightindex));
    fitposnegfract(leftindex) = (cfpos(rightindex)/cfneg(rightindex));
    fitnonegfract(leftindex) = (cfno(rightindex)/cfneg(rightindex));
    
    %calculating max values and indexes of fitted functions
    if fitval1 < cfpos(rightindex)
    fitval1 = cfpos(rightindex);
    fitidx1 = (rightindex);
    end
    if fitval2 < cfno(rightindex)
    fitval2 = cfno(rightindex);
    fitidx2 = (rightindex);
    end
    if fitval3 < cfneg(rightindex)
    fitval3 = cfneg(rightindex);
    fitidx3 = (rightindex);
    end
    
    fitradposno(leftindex) = acos(fitposnofract(leftindex));
    fitradposneg(leftindex) = acos(fitposnegfract(leftindex));
    fitradnoneg(leftindex) = acos(fitnonegfract(leftindex));
    
end

%Calculating the Verdet constant using Malus Law for Fitted functions
degchgposno = abs(fitidx1-fitidx2);
fitverdposno = abs((degchgposno*pi/180)/(blabpos*lengthwg));
fitvermalposno = abs((acos(cfpos(fitidx2)/fitval2))/(blabpos*lengthwg))

degchgposneg = abs(fitidx1-fitidx3);
fitverdposneg = 
abs((degchgposneg*pi/180)/((abs(blabneg)+abs(blabpos))*lengthwg));
fitvermalposneg = 
abs((acos(cfpos(fitidx3)/fitval3))/((abs(blabneg)+abs(blabpos))*lengthwg))

degchgnoneg = abs(fitidx2-fitidx3);
fitverdnoneg = abs((degchgnoneg*pi/180)/(blabneg*lengthwg));
fitvermalnoneg = abs((acos(cfno(fitidx3)/fitval3))/(blabneg*lengthwg))

%plot fitted functions for ratios of the data sets
cfpno = fit(lensangleseq', posnoseq','poly5','Normalize','on');
cfpneg = fit(lensangleseq', posnegseq','poly5','Normalize','on');
cfnoneg = fit(lensangleseq', nonegseq','poly5','Normalize','on');

figure;
plot(cfpno,'b',lensangleseq,posnoseq,'c*')
hold on;
plot(cfpneg,'g',lensangleseq,posnegseq,'yo')
hold on;
plot(cfnoneg,'r',lensangleseq,nonegseq,'m+')
```

```
legend('Ratio Pos I vs No I','Fitted Pos v No','Ratio Pos I vs Neg I','Fitted
Pos v Neg', 'Ratio No I vs Neg I','Fitted No v Neg','Location','NorthEast')
xlabel('degrees')
ylabel('Ratio')
```

## A.7 Simulations in Matlab comparemodel.m and Verdetfinder.m

```matlab
%% FRA Project Comparing Model

close all;

% Physical Constants
b0   = 30*1e-3;            % chosen Magnetic field intensity (T)
c    = 299792458;          % Speed of light (m/s)
q    = 1.6*10e-19;         % electron charge (Coulombs)
mu0  = 1.25663706e10-6;    % permeability of free space (m kg s-2 A-2)
e0   = 8.85418782e-12;     % permittivity of free space (m-3 kg-1 s4 A2)
Av   = 6.02e23;            % Avagadro's number

% Wave Properties
wavel      = 632e-9;          % wavelength
k          = (2*pi)/wavel;    % wave number
w          = k*c;             % omega

fhi        = pi/4;            % initial phase angle
indexsfg   = 1.6;             % for silica flint glass
vp         = w/k;             % phase velocity??? seems to be just c...

% Waveguide
rotationlab = 0;              % will be chosen
N          = 1e7; %150;       % Number of grid cells
diameter   = 0.005;           % diameter of waveguide in meters
distance   = 0.1; %1e-3;      % length of waveguide in meters
res        = distance / N;    % Resolution of grid
V          = 0;               % Verdet constant variable setup
sigma      = 10e-12;          % Conductivity; for glass can be from 10^-11 to 10^-15 (S/m)
rho        = 2203;            % Mass density kg/m^3
m          = 60080;           % Molar mass density kg/mol
volume     = m / rho;         % m^3/mol
% n         = Av/volume;      % 1/m^-3
nden = 1e+7;                  % charge density

zwg =  res:res:distance; % Coordinates of waveguide

% Time
ti         = 0;
tf         = ti + distance/c;
t          = linspace(ti, tf, N);

% Distance
di         = 0;
df         = di+distance*10e5;
d          = linspace(di, df, N);

% Source
power      = 0.001;                        % Power of laser beam (Watts)
area       = 0.5 * pi * (diameter/2)^2;    % Area of laser beam (m^2)
E0         = sqrt(2*power/(e0*c*area));    % Amplitude of electric field (V/m)
```

```matlab
            % setting up the initial Electric field
b       = repmat([0; 0; b0], 1, N);

H0   = 100;                                         % Amplitude/magnitude of 
Magnetic field
H    = H0 * exp(-1i*(w*t-k*indexsfg*zwg));          % Magnetic field equation 4
% E1  = E0 * exp(-1i*(w*t-k*indexsfg*zwg));         % Electric field equation 
4
% E1x = E0 * exp(1i*(w*t-k*indexsfg*zwg))*cos(fhi); % equation 8
% E1y = E0 * exp(1i*(w*t-k*indexsfg*zwg))*sin(fhi); % equation 8
% figure;
% plot(t,E1x)
% hold on
% plot(t,E1y)
% plot(t,E1)
%
o7      = w*H;              % delta = a cross H. What is a? related to w

% ex = ?
% ey = ?
% ez = ?
%
% D = e0*[ex 1i*o7 0; -1i*o7 ey 0; 0 0 ez]*E0; % equation 2

%Derived index of refraction
E = E0 * cos(-k*zwg(1)+fhi);            % E?
nsqrd = e0*mu0;
nfound = sqrt(nsqrd);
nsqn = E - o7;                          % n-^2 = E - delta = E - aH equation 
7
nsqp = E + o7;                          % n+^2 = E + delta = E + aH equation 
7
% np = sqrt(nsqp);
% nn = sqrt(nsqn);
np = 2.0;                               % positive index of refraction
nn = 2.0+1.4e-2;                             % (was 1.4e-7) negative index 
of refraction
%chose the above values after looking online

% Eout according to paper final derivation
% Eoutx  =  E0 * cos(k*0.5*(np-nn).*zwg+fhi).*cos(k*0.5*(np-nn).*zwg-(w*t));
% Eouty  = -E0 * sin(k*0.5*(np-nn).*zwg+fhi).*cos(k*0.5*(np-nn).*zwg-(w*t));
Eoutx  =  E0 * cos(pi*(np-nn)/wavel .* zwg - fhi) .* cos(pi*(np+nn)/wavel .* 
zwg);
Eouty  = -E0 * sin(pi*(np-nn)/wavel .* zwg - fhi) .* cos(pi*(np+nn)/wavel .* 
zwg);
Eoutz  = 0;
figure
plot(d,Eouty)
hold on
plot(d,Eoutx)
title('Linearly Polarized E field Elements Rotation Breakdown')
ylabel('E')
xlabel('Distance (meters)')
axis([0 2 -300 300])
```

```matlab
% Eout according to derivation by Matther Argall
% EoutxM  = E0 * cos(k*0.5*(np-nn).*zwg-fhi).*cos(k*0.5*(np+nn).*zwg-(w*t));
% EoutyM  = E0 * sin(k*0.5*(np-nn).*zwg-fhi).*cos(k*0.5*(np+nn).*zwg-(w*t));
% EoutzM  = 0;
% figure
% plot(t,EoutxM)
% hold on
% plot(t,EoutyM)
% title('Eout Derived MA')
%
pangle1 = atan(Eouty./Eoutx);    %equation 12
% pangle2 = pi*zwg/wavel * (np-nn)+fhi; %derived equation 12
figure
plot(d,pangle1*180.0/pi)
title('Resulting Polarization Angle')
xlabel('Distance (meters)')
ylabel('Polarization angle (degrees)')
% hold on
% plot(t,pangle2)
axis([0 2 40 55])

%Verdet Constant Calculation
gammap = (pangle1./zwg);         %verdet constant using equation 13
delta = (gammap*(wavel/pi))./b(3,:);

Vnew = diff(pangle1)./diff(zwg)/b0;

%% Verdetfinder.m %%%%%%%%%%%%%%%%%%%%%%%%%%%%%%%%%%%%%%%%%%%%%%%%%%%%%%%%%

% syms x y z

% Physical Constants
b0  = 30e-2;           % chosen Magnetic field intensity (T)
c   = 299792458;         % Speed of light (m/s)
q   = 1.6e-19;       % electron charge (Coulombs)
mu0 = 1.25663706e-6;   % permittivity of free space (m kg s-2 A-2)
e0  = 8.85418782e-12;    % permittivity of free space (m-3 kg-1 s4 A2)
Av  = 6.02e23;           % Avagadro's number

% Wave Properties
lambda      = 632e-9;          % wavelength
k           = (2*pi)/lambda;   % wave number
w           = k*c;             % omega
fhi         = 0;

% Waveguide
rotationlab  = 0;                    % will be chosen
N_lambda     = 1e6;                  % Number of wavelengths
N_per_lambda = 6;                    % Number of grid cells per wavelength
N            = N_lambda * N_per_lambda;  % Total number of grid cells
diameter     = 0.005;                % diameter of waveguide in meters
distance     = N_lambda*lambda; %1e-3;                % length of waveguide in
meters
```

```matlab
res         = distance / N;     % Resolution of grid
V           = 0;                % Verdet constant variable setup
sigma       = 10e-12;           % Conductivity; for glass can be from 10^-11 to 10^-15 (S/m)
rho         = 2203;             % Mass density kg/m^3
m           = 60080;            % Molar mass density kg/mol
volume      = m / rho;          % m^3/mol
n           = 1e-5*Av/volume;       % 1/m^-3
zwg =   res:res:distance;
% zwg       = fra_waveguide(distance,res); % Coordinates of waveguide

% Time
ti          = 0;
tf          = ti + distance/c;
t           = linspace(ti, tf, N);

% Source
power       = 0.001;                        % Power of laser beam (Watts)
area        = 0.5 * pi * (diameter/2)^2;    % Area of laser beam (m^2)
E0          = sqrt(2*power/(e0*c*area));    % Amplitude of electric field (V/m)

% setting up the initial Electric field
% b     = fra_magfield(zwg, b0);        % magnetic field along the cylinder
Ein     = zeros(3,N);
Eout    = zeros(3,N);
J       = zeros(3,N);
vd      = zeros(3,N);
E2      = zeros(3,N);
E2sum   = zeros(3,N);
b       = repmat([0; 0; b0], 1, N);
ang_pol  = zeros(1,N);
dang_pol = zeros(1,N);
Vrough = sigma*n;                       % roughly the Verdet constant
%Exo    = A*cos((w*t(1)-k*zwg(1))+fhi); % Ex orig
%Eyo    = A*cos((w*t(1)-k*zwg(1))+fhi); % Ey orig
Ein(1,1)   = E0 * cos(-k*zwg(1)+fhi);
Ein(2,1)   = E0 * cos(-k*zwg(1)+fhi);
Eout(1,1)  = Ein(1,1);
Eout(2,1)  = Ein(2,1);
J(:,1)     = sigma*Eout(:,1);                       % current density
vd(:,1)    = 1e5*J(:,1)/(n*q);                      % velocity...
%E2op      = (cross(vd,bfin))/q;
E2(:,1)    = cross(vd(:,1),b(:,1));
E2sum(:,1) = E2(:,1);
%ang_pol(1) = atan2(Eout(2,1), Eout(1,1));
%ang_pol(1) = atan(Eout(2,1) ./ Eout(1,1));

for jj=2:length(zwg);
    % Input electric field
    Ein(1,jj)  = E0 * cos(-k*zwg(jj)+fhi);
    Ein(2,jj)  = E0 * cos(-k*zwg(jj)+fhi);

    % Output electric field
    Eout(1,jj) = Ein(1,jj) + E2sum(1,jj-1);
    Eout(2,jj) = Ein(2,jj) + E2sum(2,jj-1);
    Eout(3,jj) = Eout(3,jj-1) + E2(3,1);
```

```matlab
    % Secondary electric field
    J(:,jj)  = sigma * Eout(:,jj);        % current density
    vd(:,jj) = 1e5*J(:,jj)/(n*q);         % velocity...
    E2(:,jj) = cross(vd(:,jj), b(:,jj));  % E2 = v x b
    E2sum(:,jj) = E2sum(:,jj-1) + E2(:,jj);

%     ang_pol(jj)  = atan2(Eout(2,jj), Eout(1,jj));
%     ang_pol(jj)  = atan(Eout(2,jj) ./ Eout(1,jj));
%     dang_pol(jj) = ang_pol(jj) - ang_pol(jj-1);
end

% Calculate angle of polarization
idx    = 1;
angpol = zeros(1,N/N_per_lambda);
kmax   = zeros(3,N/N_per_lambda);
for ii = 1 : N_per_lambda : N
    % Perform maximum variance analysis
    comat             = cov(Eout(:,ii:ii+N_per_lambda-1)');
    [eigvecs, eigvals] = eigs(comat);

    % Determine polarization angle
    [~, imaxvar] = max(diag(eigvals));
    kmax(:,idx)  = eigvecs(:,imaxvar);
    angpol(idx)  = atan(kmax(2,idx) / kmax(1,idx))*180/pi;

    % Next iteration
    idx = idx + 1;
end

figure;
zlambda = 1:N/N_per_lambda;
p = plot(zlambda, angpol);
ax = gca();
title('Angle of Rotation');
ax.XLabel.String = '\lambda_{N}';
ax.YLabel.String = '\theta_{Pol}';
keyboard

V = dang_pol(end)/(b(3,end).*zwg(end));
```